\documentclass[12pt,preprint]{aastex}
 
\begin{document}

\title{Compressibility and local instabilities of differentially rotating 
       magnetized gas}

\author{Alfio Bonanno}
\affil{INAF, Osservatorio Astrofisico di Catania,
       Via S.Sofia 78, 95123 Catania, Italy \\
       INFN, Sezione di Catania, Via S.Sofia 72,
       95123 Catania, Italy}


\and

\author{Vadim Urpin}
\affil{A.F.Ioffe Institute of Physics and Technology and
       Isaac Newton Institute of Chile, Branch in St. Petersburg,
       194021 St. Petersburg, Russia}


\begin{abstract}
          We study the stability of compressible cylindrical 
          differentially rotating flow in the presence of the magnetic 
          field, and show that compressibility alters qualitatively the 
          stability properties of flows. Apart from the well-known 
          magnetorotational instability that can occur even in incompressible 
          flow, there exist a new instability caused by compressibility. 
          The necessary condition of the newly found instability can 
          easily be satisfied in various flows in laboratory and 
          astrophysical conditions and reads $B_{s} B_{\phi} \Omega' 
          \neq 0$ where $B_{s}$ and $B_{\phi}$ are the radial and 
          azimuthal magnetic fields, $\Omega' =d \Omega/ds$ with $s$ 
          being the cylindrical radius. Contrary to the magnetorotational 
          instability that occurs only if $\Omega$ decreases with $s$, 
          the newly found instability operates at any sign of $\Omega'$. 
          The considered instability can arise even in a very strong 
          magnetic field that suppresses the magnetorotational 
          instability.  
\end{abstract}


\section{Introduction}

Magnetohydrodynamic instabilities and turbulence generated by these
instabilities can play an important role in enhancing transport processes 
in various astrophysical bodies, such as accretion and protoplanetary
disks, galaxies, stellar radiative zones, etc. Anomalous turbulent 
transport caused by instabilities can be particularly important in
differentially rotating and magnetized gas where a number of MHD 
instabilities may occur. It is known since the classical papers by 
Velikhov (1959) and Chandrasekhar (1960) that a differentially rotating 
flow with a negative angular velocity gradient and a weak magnetic field 
is unstable to the magnetorotational instability. This instability 
has been analyzed in detail for stellar conditions (see Fricke 1969, 
Acheson 1978, 1979). In the context of accretion disks, this instability 
was first considered by Balbus \& Hawley (1991). There are strong 
arguments that the necessary condition of the magnetorotational 
instability (a decrease of the angular velocity with cylindrical radius) 
is fulfilled in disks (see, e.g., Kaisig, Tajima, \& Lovelace 1992, 
Kumar, Coleman, \& Kley 1994, Zhang, Diamond, \& Vishniac 1994). The 
instability exists not only for short wavelength perturbations, but also 
for global modes with scales comparable to the disk height (Curry, 
Pudritz, \& Sutherland 1994, Curry \& Pudritz 1995). Likely, the 
magnetorotational instability may develop in astrophysical plasma under
a wide spectrum of conditions. The effect of radiation on stability of 
magnetized differentially rotating flow has been studied analytically by 
Blaes \& Socrates (2001) and numerically by Turner, Stone, \& Sano 
(2002) who argued that the stability criterion remains unchanged in this 
case but the growth rate can be lower. Recently, stability properties of 
magnetized accretion flows have been considered in the Kerr metric by De 
Villiers \& Hawley (2003), De Villiers, Hawley, \& Krolik (2003), and 
Hirose  et al. (2004) who found that the magnetorotational instability 
can drive turbulence in this case as well.

The numerical study of the magnetorotational instability in magnetized 
accretion disks both in local (Brandenburg et al. 1995, Hawley, Gammie 
\& Balbus 1995, Matsumoto \& Tajima 1995, Arlt \& R\"{u}diger 2001) and
global (Armitage 1998, Hawley 2000) approximations show a breakdown of
laminar Keplerian flow into well-developed turbulence. The generated 
turbulence can be an efficient mechanism of the $\alpha$-viscosity
and angular momentum transport in accretion disks (Brandenburg et al.
1996, Hawley, Gammie, \& Balbus 1996, Stone et al. 1996). At low
temperatures and high densities, however, the level of turbulence
induced by the magnetorotational instability can decrease significantly
or even turns off completely (Gammie 1996, Gammie \& Menou 1998).

In recent years, many simulations of accretion disks have been performed, 
and much of the dynamics was interpreted as a direct consequence of the
magnetorotational instability. Obviously, however, that this instability 
cannot be the only instability that operates in differentially rotating 
magnetized gaseous bodies. For example, convection may occur due to an 
outwardly decreasing entropy in radiatively inefficient magnetized 
accretion flow (Narayan et al. 2002) or in radiation-dominated 
accretion disks (see, e.g., Agol et al. (2001). If the angular velocity 
depends on the vertical coordinate, then stratification can be the 
reason of the baroclinic convection associated with non-parallel 
temperature and pressure gradients (Urpin \& Brandenburg 1998, Urpin 
2003). Stratification can also lead to a wide variety of strong 
non-axisymmetric instabilities in accretion disks (Keppens, Casse, 
\& Goedbloed 2002). 

Many previous stability analyses have adopted the Boussinesq 
approximation, and have therefore neglected the effect of compressibility.
This is valid provided that the magnetic field strength is essentially 
subthermal, and the sound speed is much greater than the Alfv\'en 
velocity, $c_{s} \gg c_{A}$. However, this can often be not the case 
in both astrophysical objects and many numerical simulations. The first 
attempt to consider the effect of compressibility on the 
magnetorotational instability was undertaken by Blaes \& Balbus (1994) in 
the context of astrophysical disks. The authors consider a very 
simplified case of the wavevector parallel to the rotation axis and 
vanishing radial magnetic field. As a result, the most interesting 
physics has been lost in their study since only the standard 
magnetorotational instability operates in this simple geometry. For 
instance, the authors concluded that the azimuthal field does not affect 
the instability criterion in all orders in $B_{\phi}$. This conclusion 
applies, however, only to the particular case considered by Blaes \& 
Balbus (1994) and is incorrect in a more general magnetic configuration.
A detailed study of the influence of $B_{\phi}$ on rotation induced 
instabilities is given by Pessah \& Psaltis (2005). These authors show that 
terms in the Lorentz force proportional to $B_{\phi}^2/s$ where $s$ is the 
cylindrical radius can be dynamically important when superthermal magnetic 
fields are considered. The curvature of toroidal field lines plays a
fundamental role in the local stability of strongly magnetized compressible
plasmas. In particular, the combined influence of compressibility and
the curvature term can give rise to two new instabilities. 

A more refined analyses of instability of a differentially rotating 
magnetized gas was undertaken by Blokland et al. (2005) and van der 
Swaluw, Blokland, \& Keppens (2005) who considered radially stratified 
disks in the cylindric limit. In particular, Blokland et al. (2005) 
investigate the influence of a toroidal field on the growth rate and 
frequency of eigenmodes and find that it leads to overstability (complex 
eigenvalue) and reduction of the growth rate. The overstable character 
of the magnetorotational instability increases as the angular velocity 
decreases. Van der Swaluw, Blokland, \& Keppens (2005) study the 
interplay between different instabilities in accretion disks and argue 
that the growth rate of convection can be essentially increased due to 
magnetorotational effects. Note that both these studies treat stability 
in the magnetic field with a vanishing radial component. 

In this paper, we show that a number of instabilities may occur in a 
compressible differentially rotating magnetized gas which are different 
from the standard magnetorotational instability. These instabilities 
appear if the magnetic field has non-vanishing radial and azimuthal 
components and can manifest themselves even if stratification is 
neglected. The considered instabilities can be oscillatory or 
non-oscillatory and arise even if the magnetic field is strong enough 
to suppress the magnetorotational instability. Our study generalizes 
the paper by Bonanno \& Urpin (2005) where shear-driven magnetic 
instabilities were considered in a compressible flow for the particular 
case when the wavevector is perpendicular to the magnetic field and the 
standard magnetorotational instability does not operate. 

The paper is organized as follows. In section 2, we discuss the main 
equations and derive the dispersion relation for a compressible 
differentially rotating magnetized gas. Stability properties of modes 
are considered in section 3, and our numerical calculations of the 
growth rate are presented in section 4. Finally, we discuss our 
conclusions in section 5.  

\section{Basic equations and dispersion relation}

We work in cylindrical coordinates ($s$, $\varphi$, $z$) with the 
unit vectors ($\vec{e}_{s}$, $\vec{e}_{\varphi}$, $\vec{e}_{z}$).
The equations of compressible MHD read  
\begin{eqnarray}
\dot{\vec{v}} + (\vec{v} \cdot \nabla) \vec{v} = - \frac{\nabla p}{\rho} 
+ \vec{g}  + \frac{1}{4 \pi \rho} (\nabla \times \vec{B}) \times \vec{B}, 
\end{eqnarray}
\begin{equation}
\dot{\rho} + \nabla \cdot (\rho \vec{v}) = 0, 
\end{equation}
\begin{equation}
\dot{p} + \vec{v} \cdot \nabla p + \gamma p \nabla \cdot 
\vec{v} = 0,
\end{equation}
\begin{equation}
\dot{\vec{B}} - \nabla \times (\vec{v} \times \vec{B}) = 0,
\end{equation}
\begin{equation}
\nabla \cdot \vec{B} = 0. 
\end{equation} 
Our notation is as follows: $\rho$ and $\vec{v}$ are the gas density and 
velocity, respectively; $\vec{B}$ is the magnetic field; $p$ is the pressure
and $\gamma$ is the adiabatic index. 

The basic state on which the stability analysis is performed
is assumed to be quasi-stationary with the angular velocity $\Omega
= \Omega(s)$ and $\vec{B} \neq 0$. We assume that gas is in hydrostatic 
equilibrium in the basic state, then
\begin{equation}
\frac{\nabla p}{\rho} = \vec{D} + \frac{1}{4 \pi \rho} 
(\nabla \times \vec{B}) \times \vec{B} \;\;, \;\;\;\;
\vec{D} = \vec{g} + \Omega^{2} \vec{s}.
\end{equation}
We consider magnetic configurations where 
both the radial and azimuthal field components are presented. The presence 
of a radial magnetic field and differential rotation in the basic state can 
lead to the development of the azimuthal field. Nevertheless, the basic 
state can be considered sometimes as quasi-stationary despite the 
development of the toroidal field. For example, if the magnetic Reynolds 
number is large, one can obtain from Eq.~(4) that the azimuthal field grows 
approximately linearly with time,
\begin{equation}
B_{\varphi}(t) = B_{\varphi}(0) + s \Omega' B_{s} t,
\end{equation} 
where $\Omega' = d \Omega/ds$, and $B_{\varphi}(0)$ is the azimuthal 
field at $t=0$. As long as the second term on the r.h.s. is small
compared to the first one or, in other words,
\begin{equation}
t < \tau_{\varphi} = \frac{1}{s \Omega'} \; 
\frac{B_{\varphi}(0)}{B_{s}},
\end{equation}
stretching of the azimuthal field does not affect significantly the
basic state; $\tau_{\varphi}$ is the characteristic timescale of 
generation of $B_{\varphi}$. As a result, the basic state can be 
treated as quasi-stationary during the time $t < \tau_{\varphi}$. If 
$B_{\varphi}(0)/ B_{s} \gg 1$, then steady-state can be maintained 
during a relatively long time before the generated azimuthal field 
begins to influence the basic state.  

We consider stability of axisymmetric short wavelength perturbations.
Small perturbations will be indicated by subscript 1, while unperturbed 
quantities will have no subscript. The linearized momentum equation is
\begin{eqnarray}
\dot{\vec{v}}_{1} + (\vec{v}_{1} \cdot \nabla) \vec{v} + (\vec{v} \cdot
\nabla) \vec{v}_{1} = - \frac{\nabla p_{1}}{\rho} + 
\nonumber \\
\frac{\rho_1}{\rho} \left[ \frac{\nabla p}{\rho} -  
\frac{1}{4 \pi \rho} (\nabla \times \vec{B}) \times \vec{B} \right] 
+ \vec{L},
\end{eqnarray}
where
$$
\vec{L} = \frac{1}{4 \pi \rho}[ (\nabla \times \vec{B}_{1}) \times \vec{B} 
+ (\nabla \times \vec{B}) \times \vec{B}_{1}]. 
$$
Taking into account that the unperturbed motion is rotation ($\vec{v} 
= s \Omega \vec{e}_{\varphi}$) and using Eq.~(6), we can transform this 
equation into 
\begin{eqnarray}
\dot{\vec{v}}_{1} \! + \! 2 \Omega \times \vec{v}_{1} \! + \! 
\vec{e}_{\varphi} s \Omega'
v_{1s}=- \frac{\nabla p_{1}}{\rho} + \frac{\rho_{1}}{\rho} \vec{D} + \vec{L}.
\end{eqnarray}
The coefficients of Eq.~(10) are not constant, but they are slowly 
varying functions of coordinates compared to perturbations. 
Therefore, one can expand coefficients in Taylor series around some radial 
point and retain only the zeroth order terms in this expansion. This is a 
good approximation as long as the wavelength of perturbations is small 
compared to the lengthscales of unperturbed quantities. Then, one can  
perform a local stability analysis of Eq.~(10) around this radius but with
constant coefficient. Since coefficients are constant, one can consider 
perturbations with the spacetime dependence $\propto \exp ( \sigma t - i 
\vec{k} \cdot \vec{r})$. where $\vec{k}= (k_{s}, 0, k_{z})$ is the wavevector. 

The term proportional to $\rho_1/\rho$ on the r.h.s. of Eq.~(10) is 
typically small in the short wavelength approximation. Indeed, we can estimate 
$p_1 \sim c_s^2 \rho_1$ where $c_s$ is the sound speed. Then, the pressure 
term in Eq.~(10) is $k_i c_s^2 (\rho_1/\rho)$ where $k_i$ is the cylindrical
component of the wavevector. Therefore, the ratio of the first two terms on 
the r.h.s. of Eq.~(10) can be estimated as $k_i c_s^2 / D_i$ where $\vec{D}=
\vec{g} + \Omega^2 \vec{s}$. In many cases of interest, the centrifugal force 
is weak compared to gravity, and we have $D_i \sim g_i$. This estimate 
applies in stellar radiative zones, solar tachocline, protostellar disks, 
galaxies, etc. However, this estimate can be incorrect in accretion disks 
with substantially superthermal magnetic fields (Pessah \& Psaltis 2005), but
we do not address these objects in the present paper. Since $c_s^2/g_i$ is 
the pressure scaleheight $H_i$, we obtained that the ratio of the first and 
second terms on the r.h.s. of Eq.~(10) is $\sim k_i H_i$. The inequality 
$k_i H_i \gg 1$ is exactly equivalent to the condition of applicability of 
the local approximation.. The terms of the order of $(k_i H_i)^{-1}$ cannot 
be taken into account consistently in the local approximation because 
inhomogeneities of the basic state produce corrections of the same order of 
magnitude. Therefore, we have to neglect the term proportional to $\vec{D}$ 
in what follows. 

Consider in more detail the Lorentz force in Eq.~(10). For the sake of
simplicity, we assume that the radial and azimuthal components the magnetic 
field in the basic state do not depend on the vertical coordinate, $\partial
B_{\varphi}/\partial z = \partial B_s/ \partial z =0$, and that $\partial 
B_z /\partial s = 0$. Then, the electric current in the basic state has only 
the vertical component, and the Lorentz force in Eq.~(10) can be represented 
as
\begin{eqnarray}
4 \pi \rho \vec{L} = \vec{e}_s  \! \left[ B_z \! \left( \! 
\frac{\partial B_{1s}}{\partial z} \! - \! \frac{\partial B_{1z}}{\partial s} 
\! \right) \! - \! B_{\varphi} 
\frac{\partial B_{1 \varphi}}{\partial s}  - \! \right.
\nonumber \\
\left. \frac{B_{1 \varphi}}{s^2} \frac{\partial (s^2 B_{\varphi})}{\partial s} 
\right] 
\! + \! \vec{e}_{\varphi} \left[ \frac{B_{1s}}{s}
\frac{\partial (s B_{\varphi})}{\partial s} \! + \! 
B_s \frac{\partial B_{1 \varphi}}{\partial s} + \right.
\nonumber \\
\left. B_{z} \frac{\partial B_{1 \varphi}}{\partial z} 
+ B_s \frac{B_{1 \varphi}}{s} \right] 
\! - \! \vec{e}_z  \left[ B_s \left( \frac{\partial B_{1s}}{\partial z}
\! - \! \frac{\partial B_{1z}}{\partial s} \right) \! \right.
\nonumber \\
\left. +  B_{\varphi} 
\frac{\partial B_{1 \varphi}}{\partial z} \right].
\end{eqnarray}
The last terms in the $s$- and $\varphi$-components of this equation are of 
the order of $(k s)^{-1}$ compared to the second terms and should be neglected
in a short wavelength approximation that does not allow to treat consistently 
corrections of such order because a slow dependence of the basic state on 
coordinates (which is neglected in a local analysis) produces corrections of 
the same order. Then, the Lorentz force with the accuracy in terms $\sim 
\lambda/s$, where $\lambda$ is the wavelength of perturbations, is given by
\begin{equation}
\vec{L} = \frac{1}{4 \pi \rho} \left[ i \vec{B} \times (\vec{k} \times
\vec{B}_1) + \vec{e}_{\varphi} \frac{B_{1s}}{s} 
\frac{\partial (s B_{\varphi})}{\partial s} \right]
\end{equation} 
The second term in this expression describes the effect caused by the presence
of electric currents in the basic state. This term can be important even in
a local approximation if the toroidal field is sufficiently strong and is 
not proportional to $s^{-1}$. 

Following similar transformations with the remaining equations in the
system given by Eq.~(1)-(5), we arrived to the linearized equations needed
for stability analysis. These equations read with accuracy in the lowest 
order in $\lambda/s$  
\begin{eqnarray}
\sigma \vec{v}_{1} + 2 \vec{\Omega} \times \vec{v}_{1}
+ \vec{e}_{\varphi} s \Omega' v_{1s}    
= \frac{i \vec{k} p_{1}}{\rho}  
\nonumber \\
- \frac{i}{4 \pi \rho} (\vec{k} \times \vec{B}_{1}) \times \vec{B} 
+ \frac{\vec{e}_{\varphi}}{4 \pi \rho} \frac{B_{1s}}{s} 
\frac{\partial (s B_{\varphi})}{\partial s} ,
\end{eqnarray}
\begin{equation}
\sigma \rho_{1} - i \rho (\vec{k} \cdot \vec{v}_{1}) = 0, 
\end{equation}
\begin{equation}
\sigma p_{1} -i \gamma p (\vec{k} \cdot \vec{v}_{1}) = 0, 
\end{equation}
\begin{equation}
\sigma  \vec{B}_{1} = \vec{e}_{\varphi} s \Omega' 
B_{1s } -i (\vec{B} \cdot \vec{k}) \vec{v}_{1} + i \vec{B} (\vec{k} 
\cdot \vec{v}_{1}), 
\end{equation}
\begin{equation}
\vec{k} \cdot \vec{B}_{1} = 0.
\end{equation}
Assuming $B_{\varphi} < (ks) B_{s}$, Eqs.~(13)-(17) may be combined after
some algebra into a sixth-order dispersion relation,
\begin{equation}
\sigma^{6} + a_{4} \sigma^{4} + a_{3} \sigma^{3} + a_{2} \sigma^{2} 
+ a_{1} \sigma + a_{0}= 0. 
\end{equation}
The coefficients of this equation can be expressed in terms of 
characteristic frequences. We have with accuracy in the lowest order in 
$\sim (ks)^{-1}$  
\begin{eqnarray}
a_{4} = \omega^{2}_{0} + \omega^{2}_{A} + \Omega^{2}_{e}, \;\;
a_{3} = \omega^{3}_{B \Omega} + 2 i \Omega \Omega_{A \varphi} \omega_A, 
\nonumber \\
a_{2} = \omega^{2}_{A} (\omega^{2}_{s} + \omega^{2}_{0}) +
\Omega^{2}_{e} (\mu \omega^{2}_{0} + \omega_{A} k_{s} c_{As}) \nonumber \\
- 4 \Omega^{2} \omega_{A} k_{z} c_{Az} + i \Omega_{A \varphi} \omega_{A}^2
c_{A \varphi} k_s, \nonumber \\
a_{1} = \omega^{3}_{B \Omega} (\omega^{2}_{A} + \mu \Omega^{2}_{e})
+ 2 i \Omega \Omega_{A \varphi} \omega_A [ \mu \omega_0^2 + 
\nonumber \\
\omega_A (k_s c_{A s} - k_z c_{A z})], \nonumber \\
a_{0} = (\omega^{2}_{s} \omega^{2}_{A} + i \Omega_{A \varphi} \omega_A
c_{A s} c_{A \varphi} k^2 )( \omega^{2}_{A} +
\mu \Omega^{2}_{sh}), \nonumber
\end{eqnarray}
where $\mu = k^{2}_{z}/k^{2}$. The characteristic frequencies are given by
\begin{eqnarray}
\Omega^{2}_{e} = 4 \Omega^{2} + \Omega^{2}_{sh} \;, \;\;\;
\Omega^{2}_{sh} = 2 s \Omega \Omega' \;, \nonumber \\
\omega^{2}_{0}= \omega^{2}_{s} + \omega^{2}_{m} , \;\; 
\omega^{2}_{s} = c^{2}_{s} k^{2} , \;\; \omega^{2}_{m} = c^{2}_{A} 
k^{2} \;, \nonumber \\
\omega_{A} = \vec{k} \cdot \vec{c}_{A} \;, \;\;\; \vec{c}_{A} =
\vec{B} /\sqrt{4 \pi \rho} \;, \nonumber \\
\omega^{3}_{B \Omega} = k^{2} c_{A \varphi} c_{A s} s \Omega' , \;\;
\Omega_{A \varphi} = \frac{c_{A \varphi}}{s} \! \left( \! 1 \! + \!
\frac{s}{B_{\varphi}}
\frac{\partial B_{\varphi}}{\partial s} \! \right)
\nonumber,
\end{eqnarray}
where $c_{s} = \sqrt{\gamma p/ \rho}$ is the sound speed. Eq.~(18) 
describes six modes that can generally exist in a compressible
rotating magnetized gas. 

Before to solve Eq.~(18), we consider several particular cases and 
compare our dispersion relation with those derived previously by other 
authors. In the incompressible limit when the sound speed $c_{s}$ and, 
hence, the sound frequency $\omega_{s}$ are very large, we obtain from 
Eq.~(18) the dispersion relation for the standard magnetorotational 
instability,
\begin{equation}
\sigma^{4} + \sigma^{2} (2 \omega^{2}_{A} + \mu \Omega^{2}_{e}) +
\omega^{2}_{A} (\omega^{2}_{A} + \mu \Omega^{2}_{sh}) =0
\end{equation}
(Velikhov 1959). This equation allows unstable solutions if
\begin{equation}
\mu \Omega^{2}_{sh} < - \omega^{2}_{A},
\end{equation}
that represents the criterion of the magnetorotational instability in the
incompressible limit. It is seen that the magnetorotational instability 
can be entirely suppressed if the magnetic field is sufficiently strong 
(see, e.g., Urpin 1996, Kitchatinov \& R\'{u}diger 1997, Balbus \& Hawley 
1998).  

In the particular case $k_{s}=B_{s}=0$ and $\partial (s B_{\varphi})/
\partial s =0$ (no currents in the basic state), we obtain from Eq.~(18)
the dispersion relation derived by Blaes \& Balbus (1994) in their 
analysis of the effect of compressibility,
\begin{equation}
\sigma^{6} + q_{4} \sigma^{4} + q_{2} \sigma^{2} + q_{0} = 0,
\end{equation}
where
\begin{eqnarray}
q_{4}= k^{2} ( 2 c^{2}_{Az} + c^{2}_{A \phi} + c^{2}_{s}) + 
\Omega^{2}_{e}, \nonumber \\
q_{2} = k^{2} [ k^{2} c^{2}_{Az} (c^{2}_{A} +
2 c^{2}_{s} ) + \Omega^{2}_{e} (c^{2}_{s} + c^{2}_{A \phi}) +
c^{2}_{Az} \Omega^{2}_{sh}], \nonumber \\
q_{0} = k^{4} c^{2}_{s} c^{2}_{Az} (k^{2} c^{2}_{Az} + 
\Omega^{2}_{sh}). 
\nonumber
\end{eqnarray}
From Eq.~(21), Blaes \& Balbus (1994) deduced that instability of 
compressible and incompressible gas is determined by the same criterion (20).

Note that Eq.~(18) in the case $B_{s}=0$ differs from the dispersion relation 
derived by Pessah \& Psaltis (2005) (see Eq.~(25) of their paper). These
authors consider the influence of the ``curvature terms'' on stability of 
differential rotation in accretion disks. Likely, the difference is caused 
mainly by different assumptions regarding the basic state. Pessah \& Psaltis 
(2005) derive the dispersion relation for the midplane of accretion disks 
and a substantially superthermal magnetic field. As it was mentioned above, 
our study is not addressed such objects.

Eq.~(18) is very complicated for analysis, and we consider in this paper 
only the case when the effect of electric currents can be neglected. Certainly,
this is valid if $B_{\varphi} \propto 1/s$, but the contribution of currents
can be small in some other cases as well. Assuming $s \Omega' \sim \Omega$, 
we can estimate that the terms caused by the electric current ($\propto 
\Omega_{A \varphi}$) in the coefficients $a_3$ and $a_2$ provide only small 
corrections of the order of $\lambda/s$ to $\omega_{B \Omega}^3$ and 
$\omega_0^2 \omega_{A}^2$, respectively, and can be neglected in these 
coefficients. The effect of currents in $a_1$ is unimportant is 
\begin{equation}
B_{\varphi}^2 \ll (ks) B_s^2.
\end{equation}   
Finally, the influence of currents in the coefficient $a_0$ is negligible if
\begin{equation}
\frac{c_A^2}{c_s^2} \ll ks,
\end{equation}
that is fulfilled usually in astrophysical conditions. Conditions (22) and 
(23) determine the domain of parameters where the effect of electric currents 
on stability properties can be neglected. For this reason, we consider in 
the present paper stability in the domain given by Eqs.~(22) and (23) and 
will suppose $\Omega_{A \varphi} =0$ in Eq.~(18).    

It was shown by Bonanno \& Urpin (2006) that Eq.~(18) at $\Omega_{A \varphi}
=0$ has unstable solutions
even if the criteria of the magnetorotational and Rayleigh instability are
not satisfied. The authors consider stability of perturbations with 
perpendicular $\vec{k}$ and $\vec{B}$. The dispersion relation reads in
this case 
\begin{equation}
\sigma^{5} + \sigma^{3} (\omega^{2}_{0} + \Omega^{2}_{e})
+ \sigma^{2} \omega^{3}_{B \Omega} + \sigma \mu \Omega^{2}_{e} 
\omega^{2}_{0}
+ \mu \Omega^{2}_{e} \omega^{3}_{B \Omega} = 0. 
\end{equation}
As it is seen from Eq.~(19), perturbations with $\vec{k} \cdot \vec{B} =0$ 
are not subject to the magnetorotational instability in the incompressible 
limit. Nevertheless, as it was shown by Bonanno \& Urpin (2006), Eq.~(24) 
can have unstable solutions if $\omega_{B \Omega} \neq 0$. 
Depending on parameters, the roots of Eq.~(24) can vary within a wide range,
but the growth rate of unstable modes is always proportional to a power 
of $\omega_{B \Omega}$ (if the magnetorotational and Rayleigh instability
do not occur in a flow). 

The presence of terms $\propto \omega_{B \Omega}$ in Eq.~(18) is always 
crucial for instability. To illustrate this point, we consider the solution 
of dispersion relation (18) in the case of a non-vanishing Alfven frequency, 
$\omega_A \neq 0$. For the sake of simplicity, we assume that $\Omega_e^2 = 
0$ ($\Omega \propto s^{-2}$) and consider the case of a cold gas with small
$c_s$. Since the magnetorotational instability does not occur if 
$c_A \gg c_s$, the flow is not subject to any known rotation-induced 
instabilities. Dispersion relation (18) reads in this case
\begin{eqnarray}
\sigma^{5} + \sigma^{3} (\omega^{2}_{m} + \omega^{2}_{A})
+ \sigma^{2} \omega^{3}_{B \Omega} + \sigma [\omega^{2}_{A} \omega^2_{m}  
\nonumber \\
- 4 \Omega^2 \omega_{A} k_z c_{Az}]
+ \omega^{2}_{A} \omega^{3}_{B \Omega} = 0. 
\end{eqnarray}
We consider stability of a special type of perturbation with the wavevector, 
satisfying the condition $k^2 = 4 h \Omega^2/ c_{A}^2$ where $h = k_z c_{Az}
/ \omega_{A}$. If $s \Omega/ c_{A} \gg 1$, our choice of $k$ is compatible 
with a short wavelength approximation since it implies $ks \gg 1$. For such 
perturbations, we have from Eq.~(25)    
\begin{equation}
\sigma^{3} (\sigma^{2} + \omega^{2}_{m} + \omega^{2}_{A})
+ (\sigma^{2} + \omega^{2}_{A} ) \omega^{3}_{B \Omega} = 0. 
\end{equation}
If the radial component of the magnetic field is small, $B_s \ll B_z \sim 
B_{\varphi}$, then the frequency $\omega_{B \Omega}$ can be considered as a 
small perturbation in Eq.~(26), and we can solve this equation by making use 
of the perturbation procedure. Two roots are non-vanishing even in the zeroth
order, and they can be obtained approximately by neglecting $\omega_{B \Omega}$
in Eq.~(26). Then, we have
\begin{equation}
\sigma_{1,2} 
\approx \pm i \sqrt{\omega_{m}^2 + \omega_{A}^2}.
\end{equation}
The growth rate of oscillatory modes is small compared to the frequency in 
this case, and we neglect it in the expression for $\sigma_{1,2}$. Three 
other roots are proportional $\omega_{B \Omega}^3$, and their frequencies 
are small compared to $\omega_m$ and $\omega_A$. Therefore, calculating these 
roots, we can neglect $\sigma^2$ compared to $\omega_m^2$ and $\omega_A^2$ in 
both brackets in Eq.~(26). Then, we obtain
\begin{equation}
\sigma_{3, 4, 5}^3 \approx - \frac{\omega_{A}^2  
\omega_{B \Omega}^3}{\omega_{A}^2 + \omega_{m}^2} \;, \;\; 
\end{equation}
Since $\omega_{A} \sim \omega_{m}$, the growth rate of unstable modes can be 
estimated as Re $\sigma \sim |\omega_{B \Omega}|$. Taking into account that 
$k^2 \sim \Omega^2 / c_{A}^2$, we obtain 
\begin{equation}
{\rm Re} \; \sigma \sim \Omega \;\left( \frac{B_s}{B_{\varphi}} \right)^{1/3}
\end{equation}
(we assume that $s \Omega' \sim \Omega$). Note that similar solutions 
with Re $\sigma \sim |\omega_{B \Omega}|$ exist also in flows with non-zero 
sound speed if $c_s$ satisfies the condition  $c_s \ll \min(c_{As}, \Omega/k)$
where $c_{As}= B_s/ \sqrt{4 \pi \rho}$.

\section{Criteria of instability}

The conditions under which Eq.~(18) has unstable solutions with a positive
real part can be obtained by making use of the Rooth-Hurwitz theorem 
(see Henrici 1977, Aleksandrov, Kolmogorov, \& Laurentiev 1985, Toth, 
Szili, \& Zach\'ar 1998). The Rooth-Hurwitz criteria for a sixth-order 
equation are rather complex in the general case, but they are very much 
simplified for Eq.~(18) where the coefficient before $\sigma^{5}$ is 
vanishing. In accordance with these criteria, Eq.~(18) has unstable 
solutions if one of the following inequalities is fulfilled
\begin{eqnarray}
a_{3} > 0 \;,  \\
a^{2}_{3} > 0 \, \\
a_{1} (a_{3} a_{4} - a_{1}) - a_{2} a^{2}_{3} < 0 \;, \\
a^{2}_{1} (a_{3} a_{4} - a_{1}) + a^{2}_{3} (a_{0} a_{3} -
a_{1} a_{2}) < 0 \;, \\
a_{0} < 0 \;.  
\end{eqnarray}
One of conditions (30) and (31) is always fulfilled if $a_{3} \neq 0$, or
\begin{equation}
\omega^{3}_{B \Omega} \neq 0.
\end{equation}
Apart from differential rotation, this criterion requires non-vanishing 
radial and toroidal components of the magnetic field. Condition (35) can 
generally be satisfied for both the inward and outward decreasing angular 
velocity in contrast to the criterion of the magnetorotational 
instability (19) that is fulfilled only if $\Omega' <0$.

Since Eq.~(18) has always unstable solutions if $\omega_{B \Omega} \neq 
0$, remaining criteria (32)-(35) can be treated only in the case 
$\omega_{B \Omega} = 0$. In this case, however, we have $a_{1} = a_{3} 
= 0$, and neither Eq.~(32) nor Eq.~(33) is fulfilled. Therefore, these 
criteria do not provide any new conditions of instability. Eq.~(34) 
yields the standard criterion of the magnetorotational instability (19) 
which does not overlap condition (35). For instance, the 
magnetorotational instability can occur even if $B_{s}=0$ or $B_{\phi}
=0$, when the magnetic shear-driven instability (35) does not operate. 
On the contrary, the shear-driven instability can arise under conditions 
when the magnetorotational instability does not occur. For example, as 
it was mentioned, the magnetorotational instability is suppressed by a 
sufficiently strong magnetic field (see criterion (19)) but instability 
(35) does not exhibit such suppression and may occur even if the 
magnetic field is relatively strong (see Eq.~(23)).  

It should be emphasized that differential rotation always generates the 
azimuthal magnetic field from the radial one and, in fact, the necessary 
and sufficient condition of instability (35) is a non-vanishing radial 
magnetic field. Therefore, we have to conclude that all differentially 
rotating gaseous flows with a non-vanishing radial magnetic field are 
unstable, and this conclusion does not depend on the sign of $B_{s}$ or 
$\Omega'$. 

Note that there is no one-to-one correspondence between the roots
and criteria. In some cases, few modes are unstable if only one criterion 
is fulfilled and, on the contrary, only one mode can be unstable if
few criteria are satisfied. Generally, Eq.~(18) describes two Alfven
waves, two slow and two fast magnetosonic waves, however, it can be
difficult to discriminate between these modes if $\Omega \sim 
\omega_{s} \sim \omega_{A}$. To demonstrate which modes are unstable
if condition (35) is satisfied we consider a very slow rotating gas 
with $\omega_{s} \sim \omega_{A} \gg \Omega \sim s \Omega'$. In this 
case, we can restrict the solution of Eq.~(18) only by terms of the 
zeros and first orders in $\Omega$ and $\Omega'$. Then, Eq.~(18) yields
for such slow rotation
\begin{equation}
(\sigma^{2} + \omega_{A}^{2})[(\sigma^{4} + \sigma^{2} \omega_{0}^{2}
+ \omega_{s}^{2} \omega_{A}^{2}) + \sigma \omega_{B \Omega}^{3}]=0.
\end{equation} 
The term $\propto \omega_{B \Omega}^3$ describes a small linear corrections 
associated to differential rotation. The magnetorotational instability does 
not appear in the linear approximation in $\Omega$ since it is caused by 
quadratic in the angular velocity terms. A couple of roots of Eq.~(36) 
corresponds to to the Alfven waves, $\sigma = \pm i \omega_{A}$, which are 
stable in the linear approximation even if criterion (35) is fulfilled. Note,
however, that Alfven waves can be unstable if non-linear in $\Omega$
terms are taken into account. Fast and slow magnetosonic modes satisfy 
the dispersion relation of the fourth order,  
\begin{equation}
\sigma^{4} + \sigma^{2} \omega_{0}^{2} + \omega_{s}^{2} \omega_{A}^{2} 
= - \sigma \omega_{B \Omega}^{3}.
\end{equation} 
This equation can be solved by making use of the perturbation procedure
since $\omega_{B \Omega}^{3}$ is small. We can represent $\sigma$ as
$\sigma_{0} + \Delta \sigma$ where $\sigma_{0}$ is the solution of the
dispersion equation at $\Omega=0$ and $\Delta \sigma$ is small correction
caused by slow rotation. Then, it can be obtained from Eq.~(37) that
\begin{equation}
\sigma_{0}^{2} = - \frac{\omega_{0}^{2}}{2} \pm 
\sqrt{ \frac{\omega_{0}^{4}}{4} - \omega_{s}^{2} \omega_{A}^{2}},
\end{equation}
and 
\begin{equation}
2 \Delta \sigma = \mp \omega_{B \Omega}^{3} ( \omega_{0}^{4}
- 4 \omega_{s}^{2} \omega_{A}^{2} )^{-1/2}.
\end{equation}
The imaginary part of $\sigma$ is determined by $\sigma_{0}$ whereas the
real part which is responsible for instability is determined by $\Delta
\sigma$. The upper and low sign in Eq.~(39) corresponds to the slow and
fast magnetosonic modes, respectively. Only these modes can be unstable 
under condition (35) in the linear in $\Omega$ approximation.
If $\omega_{B \Omega}^{3} < 0$, then the slow mode is unstable and, on 
the contrary, if $\omega_{B \Omega}^{3} >0 $ then the fast mode is 
unstable. Note, however, that this simple picture can be changed if 
rotation is fast, $\Omega \sim \omega_{s} \sim \omega_{A}$. The Alfven 
mode can become unstable in this case as well.

Our conclusion regarding instability caused by a combined effect of
compressibility and differential rotation in a magnetized gas appears
to be in contrast to the result obtained in the review by Balbus \& 
Hawley (1998). These authors claim that in the presence of Keplerian 
differential rotation, the fast magnetosonic mode remains stable, and 
only slow magnetosonic mode can be unstable. In fact, however, there 
is no contradiction between these conclusions because Balbus \& Hawley 
(1998) consider the basic magnetic configurations with only vertical 
and azimuthal magnetic field components, but the radial field is 
vanishing in their model. As it follows from our analysis, 
$\omega_{B \Omega}=0$ in this case, and the instability associated to 
compressibility should not occur in the model considered by Balbus \& 
Hawley (1998).

\section{Numerical results}

To calculate the growth rate of instability, it is convenient to 
introduce dimensionless quantities
\begin{eqnarray}
\Gamma= \frac{\sigma}{\Omega_{e}} \;,\;\; 
q = \frac{4 \Omega^{2}}{\Omega^{2}_{e}} \;,\;\;
f_{s} = \frac{\omega^{2}_{s}}{\Omega^{2}_{e}} \frac{1}{x^{2}} \;,\;\;
f_{m} = \frac{\omega^{2}_{m}}{\Omega^{2}_{e}} \frac{1}{x^{2}} \;,
\nonumber \\
\varepsilon = \frac{(\vec{k} \cdot \vec{B})^{2}}{k^{2} B^{2}} 
\;,\;\;
\delta = \frac{c_{A \phi} c_{A s}}{c^{2}_{A}} \;,\;\;
h = \frac{k_{z} c_{Az}}{\omega_{A}} \;,\;\; x =ks \nonumber 
\end{eqnarray}
(we assume $\Omega^{2}_{e} >0$). Note that the sign of $\Omega'$ is 
determined by the parameter $q$: $\Omega'$ is positive if $q<1$ and 
negative if $q>1$. Then, Eq.~(18) transforms into
\begin{eqnarray}
\Gamma^{6} + \Gamma^{4} [1 + f_{s} x^{2} + f_{m} x^{2} (1 + \varepsilon)] 
+ \Gamma^{3} \; \frac{1-q}{\sqrt{q}} \delta f_{m} x^{2}  \nonumber \\
+ \Gamma^{2} x^{2} [ \varepsilon f_{m} (2 f_{s} x^{2} + f_{m} x^{2} + 1 - 
h -hq)  \nonumber \\
+ \mu (f_{s} + f_{m})] + 
\Gamma \; \frac{1-q}{\sqrt{q}} (\mu + \varepsilon f_{m} x^{2})
\delta f_{m} x^{2}  \nonumber \\
+ \varepsilon f_{s} f_{m} x^{4}[ \varepsilon f_{m} x^{2} + 
\mu (1-q)] =0.
\end{eqnarray}
The dependence on the wavelength is characterized by the parameter $x^{2}$
in this equation. Eq.~(40) was solved numerically for different values of 
the parameters by computing the eigenvalues of the matrix whose 
characteristic polynomial is given by Eq.~(40) (see Press et al. 1992 for 
details).

In Fig.~1, we plot the dependence of the real part of $\Gamma$ on $f_{m}$ 
for the flow with $q=4$ ($\Omega \propto s^{-3/2}$) and $\mu = \varepsilon
=0.3$, $\delta=0.2$, $h=0.5$ and $x^{2} =10$. Solid and dashed line 
correspond to complex and real roots, respectively. We choose the value 
$f_{s} =10$ for Fig.~1. Therefore, $f_{s} \gg f_{m}$ for the considered 
range of $f_{m}$ and, hence, $c_{s} \gg c_{A}$ that corresponds to the 
case when compressibility is almost negligible. The profile of $\Omega$ 
with $q=4$ is unstable to the magnetorotational instability, and our 
calculations show that this instability can actually occur if $f_{m}$ is 
not large. Criterion of the magnetorotational instability (19) can be 
rewritten in terms of the dimensionless parameters as 
\begin{equation}
f_{m} < 3 \mu /\varepsilon  x^{2}.
\end{equation}
For given values of $\mu$, $\varepsilon$, and $x^{2}$, we obtain that this
inequality satisfies in the region where $f_{m} < 0.3$. It is seen from 
Fig.~1 that there is a real root with large Re $\Gamma$ ($\sim 0.1-0.4$) 
that can be identified with the magnetorotational instability (which is
always non-oscillatory in the incompressible limit). Apart from this root,
however, there exist another unstable root which is much smaller for the
chosen values of parameters. The new root is positive in the region 
$f_{m} > 0.3$ where the magnetorotational instability is completely
suppressed. The new instability exists even in the region of very 
large $f_{m}$ in complete agreement with criterion (32). The important 
point is that this unstable root is caused by compressibility and does 
not appear in the Boussinesq approximation. The growth rate of the new 
instability is  relatively small (Re $\Gamma \sim 0.01$) for the chosen 
parameters because $c_{s} \gg c_{A}$, and departures from the 
incompressible limit are small. The situation, however, can change 
qualitatively if the magnetic field is stronger, and we are beyond the 
necessary condition of the Boussinesq approximation, $c_{A} \ll c_{s}$.


\begin{figure}
\plotone{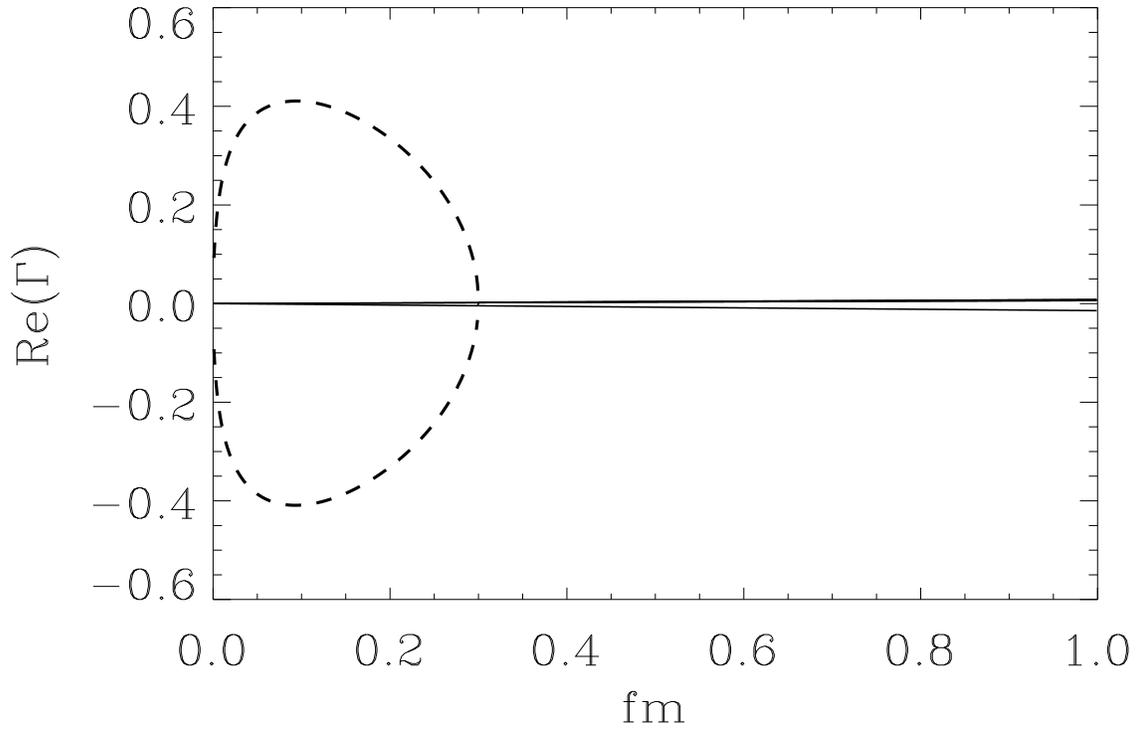}
\caption{The dependence of the real part of $\Gamma$ on $f_{m}$ for $\mu
=\varepsilon= 0.3$, $\delta=0.3$, $h=0.5$, $x^{2} = 10$, and $f_{s}=10$.
The angular velosity profile is given by $\Omega \propto s^{-3/2}$.}
\end{figure}

In Fig.~2, we show the same dependence as in Fig.~1 but for the case 
$f_{s}=0.3$ (other parameters are same) when compressibility plays much 
more important role. The Boussinesq approximation does not apply in this 
case except the region of small $f_{m} \ll f_{s}$. Nevertheless, the 
magnetorotational instability can still occur in the region $f_{m} < 
0.3$, and its growth rate is in a good agreement with the incompressible 
case. In the region of a strong field, however, there exists a new 
instability caused by shear and the magnetic field. Compressibility is 
crucial for this new instability, and it cannot be obtained in the 
incompressible limit. The growth rate of the newly found  instability is 
rather high, Re $\Gamma \sim 0.1-0.2$, in the region $f_{m} > 0.3$ where 
the magnetorotational instability is suppressed. At a fixed wavelength 
of perturbations, the growth rate increases slightly if the magnetic 
field becomes stronger. The new instability is not suppressed even by a 
very strong magnetic field. Calculations show that further decrease of 
$f_{s}$ (that is equivalent to increase of compressibility) leads to a 
higher growth rate that can be even as high as Re $\Gamma \sim 0.5$ in 
some cases. 

\begin{figure}
\plotone{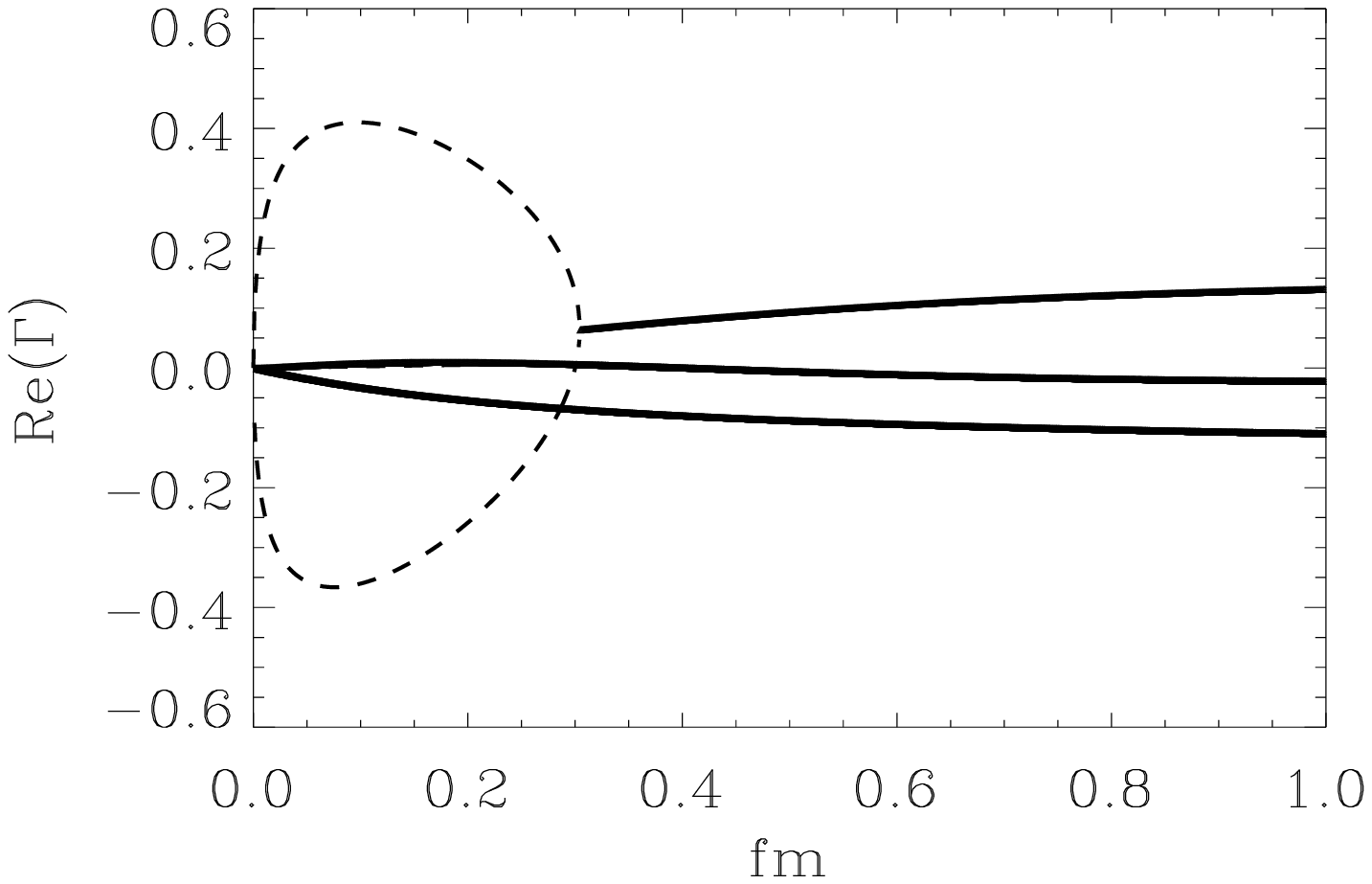}
\caption{Same as in Fig.~1 but for $f_{s}=0.3$. Solid and dashed lines 
correspond to complex and real roots, respectively.}
\end{figure}

The number of unstable roots and the growth rate are sensitive to a 
relative orientation of the azimuthal and radial magnetic fields. This 
orientation is characterized by the parameter $\delta$, and we plot the 
growth rate versus $f_{m}$ for $\delta=-0.2$ in Fig.~3. In contrast to 
Fig.~2, there are two couples of unstable complex conjugate roots for 
$\delta=-0.2$. As it was mentioned in Sec.3, apart from magnetosonic 
waves the Alfven mode can be unstable as well if rotation is sufficiently
fast. One couple has a larger growth rate (Re $\Gamma \sim 0.1$) whereas 
another couple grows substantially slower (Re $\Gamma \sim 0.05$). As 
usual, compressible instabilities are most efficient in the region where 
the magnetorotational instability does not occur ($f_{m} > 0.3$). For 
$\delta=-0.2$, however, the compressible instabilities can arise even at 
$f_{m} < 0.3$, where they work in parallel to the magnetorotational 
instability. 

\begin{figure}
\plotone{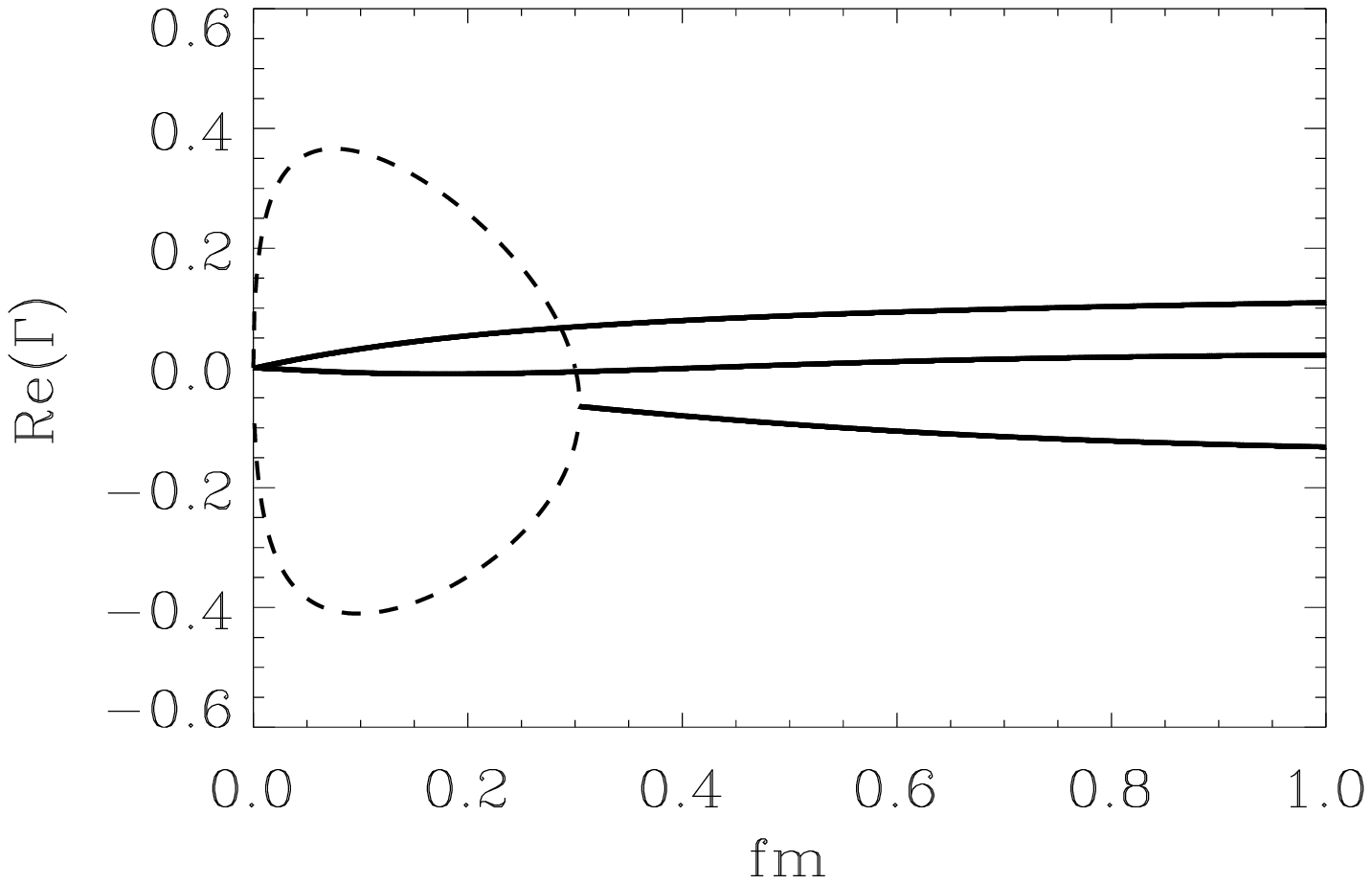}
\caption{Same as in Fig.~2 but for $\delta=-0.2$.}
\end{figure}

The above three figures illustrate the behavior of roots in the case 
when $\Omega' < 0$ and, hence, the magnetorotational instability can  
operate if the magnetic field satisfies Eq.~(41). The newly found 
instability, however, can arise in the case $\Omega' > 0$ as well. In 
Fig.~4, we plot the real parts of roots for $q=0.5$ that corresponds to 
$\Omega \propto s^{2}$ (other parameters are same as in Fig.~1). It is 
seen from the figure that there is no large root that could correspond 
to the magnetorotational instability in the region of low magnetic 
fields. Instead, there are three couples of complex conjugate roots, 
and one of these couples has positive Re $\Gamma$. These roots represent 
a relatively weak shear-driven instability that operates at any field 
strength. Obviously, the instability is weak (Re $\Gamma \sim 0.01$) 
because $c_{s} \gg c_{A}$ (or $f_{s} \gg f_{m}$) for the chosen 
parameters. However, the growth rate becomes larger as the magnetic 
field increases and departures grow from the incompressible limit.

\begin{figure}
\plotone{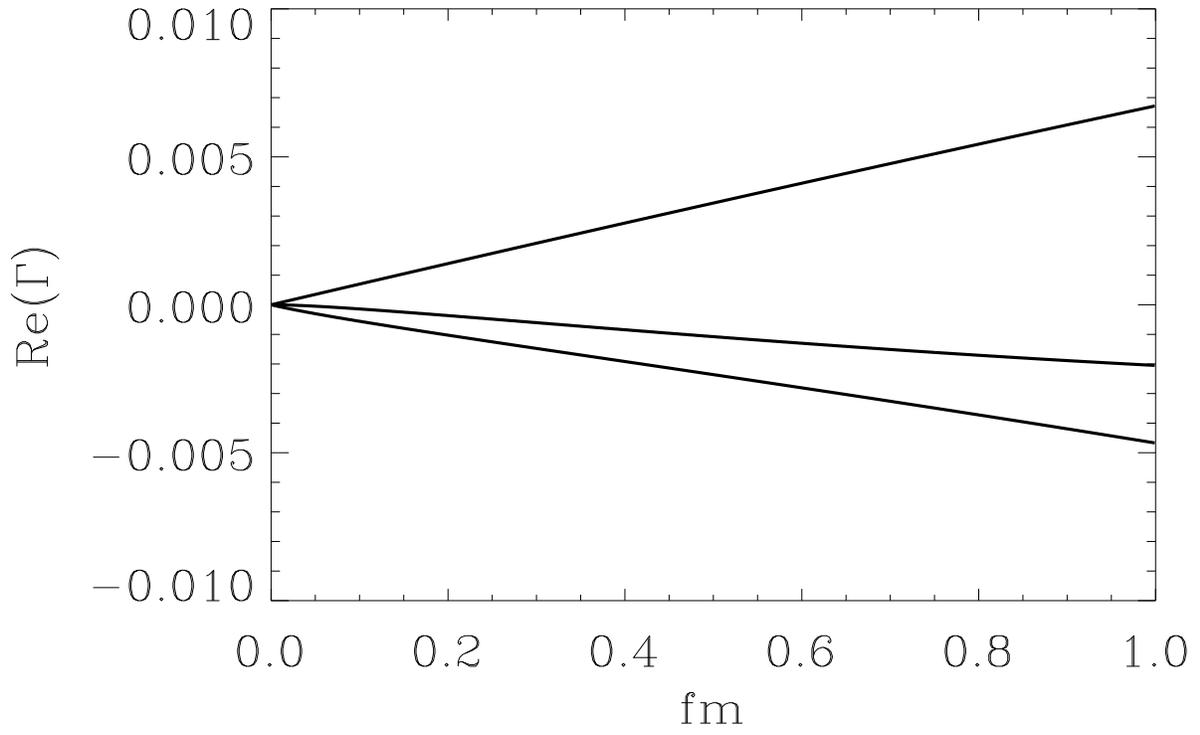}
\caption{Same as in Fig.~1 but for the case $q=0.5$ when the 
magnetorotational instability does not occur.}
\end{figure}

In Fig.~5, we show the growth rate for the case when the magnetorotational
instability still does not occur ($q=0.5$ and $\Omega \propto s^{2}$) but 
the effect of compressibility is more significant than in Fig.~4 since 
$f_{s}=0.3$. Again, all modes are complex, and two couples of complex
roots are unstable. Due to compressibility, the growth rate of the most 
unstable modes can reach rather high values $\sim 0.04-0.06 \Omega_{e}$. 
Since the epicyclic frequency is large, $\Omega^{2}_{e} = 8 \Omega^{2}$, 
we estimate the growth rate as $0.1-0.2 \Omega$, that implies that the 
growth time of instability is of the order of the rotation period, $P = 2 
\pi /\Omega$, in this case. Therefore, the considered instability turns 
out to be very rapid even in those cases when the magnetorotational 
instability cannot operate. Note that the newly found instability may
occur even if the magnetic field is very strong and the magnetic pressure 
is greater than the gas pressure (or $f_{m} > f_{s}$).

\begin{figure}
\plotone{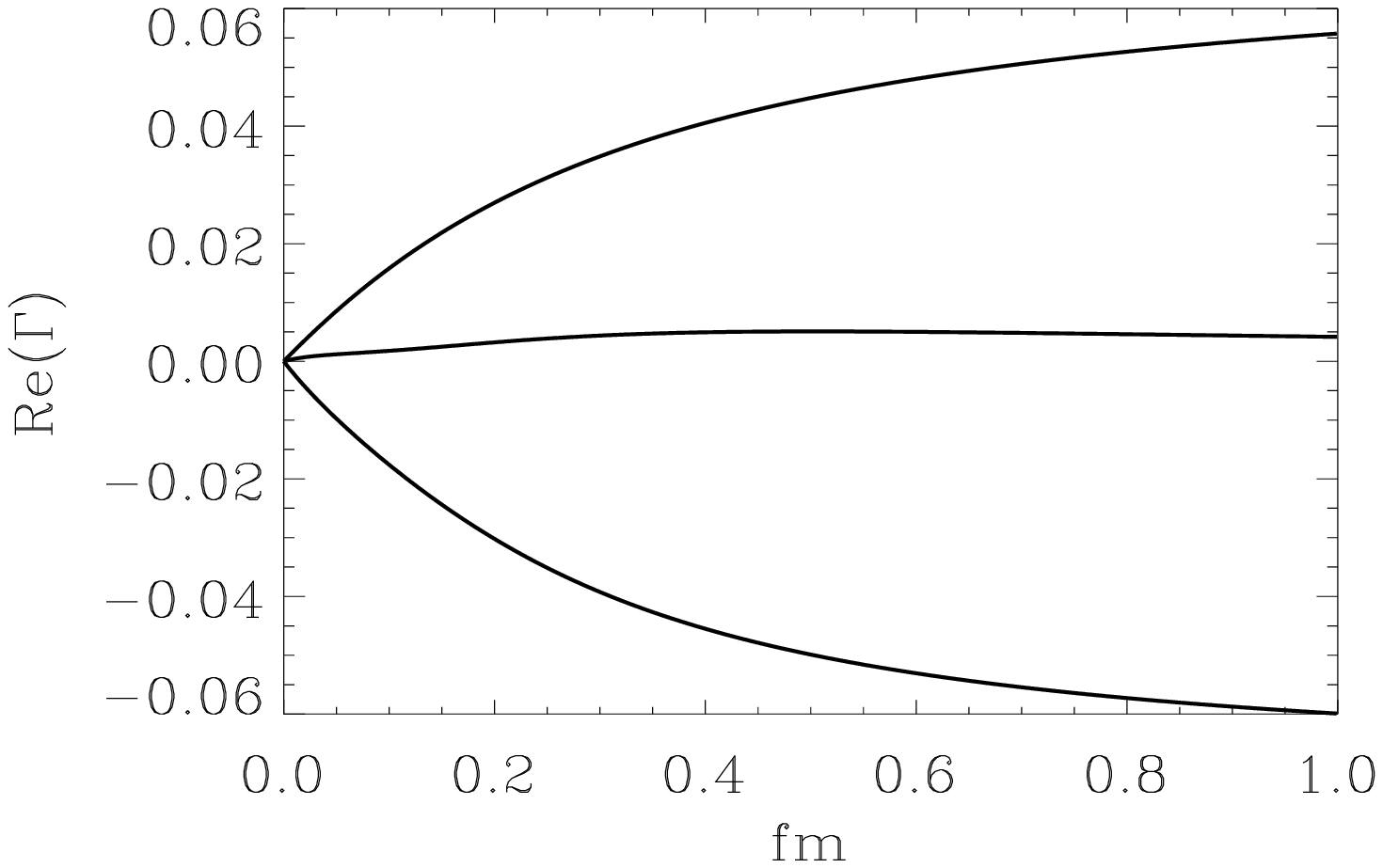}
\caption{Same as in Fig.~4 but for $f_{s}=0.3$.}
\end{figure}

Generally, a dependence of the growth rate on the parameters $\mu$ and
$\varepsilon$ is relatively weak whereas the value $h$ plays more 
important role for instability. In Fig.~6, this point is illustrated by 
calculations of Re $\Gamma$ for $h=2$ that is larger than in the previous 
figures. In fact, the parameter $h$ characterizes the direction of 
$\vec{k}$ with respect to the magnetic field. The value $h=2$ implies 
that $k_z/k_s = - 2 B_s/B_z$. Calculations have been performed for 
$f_{s}=0.3$ to compare with the previous figure. A higher value of $h$ 
does not change complexity of the modes, and all roots are complex. 
Generally, an increase in $h$ can lead to a larger number of unstable 
modes and a higher growth rate as it is seen from the figure. The 
dependence of Re $\Gamma$ on the field strength becomes non-monotonic 
with maximum at $f_{m} \sim f_{s}$. The maximum growth rate in Fig.~6 
is approximately 3 times greater than in the case $h=0.5$. 

\begin{figure}
\plotone{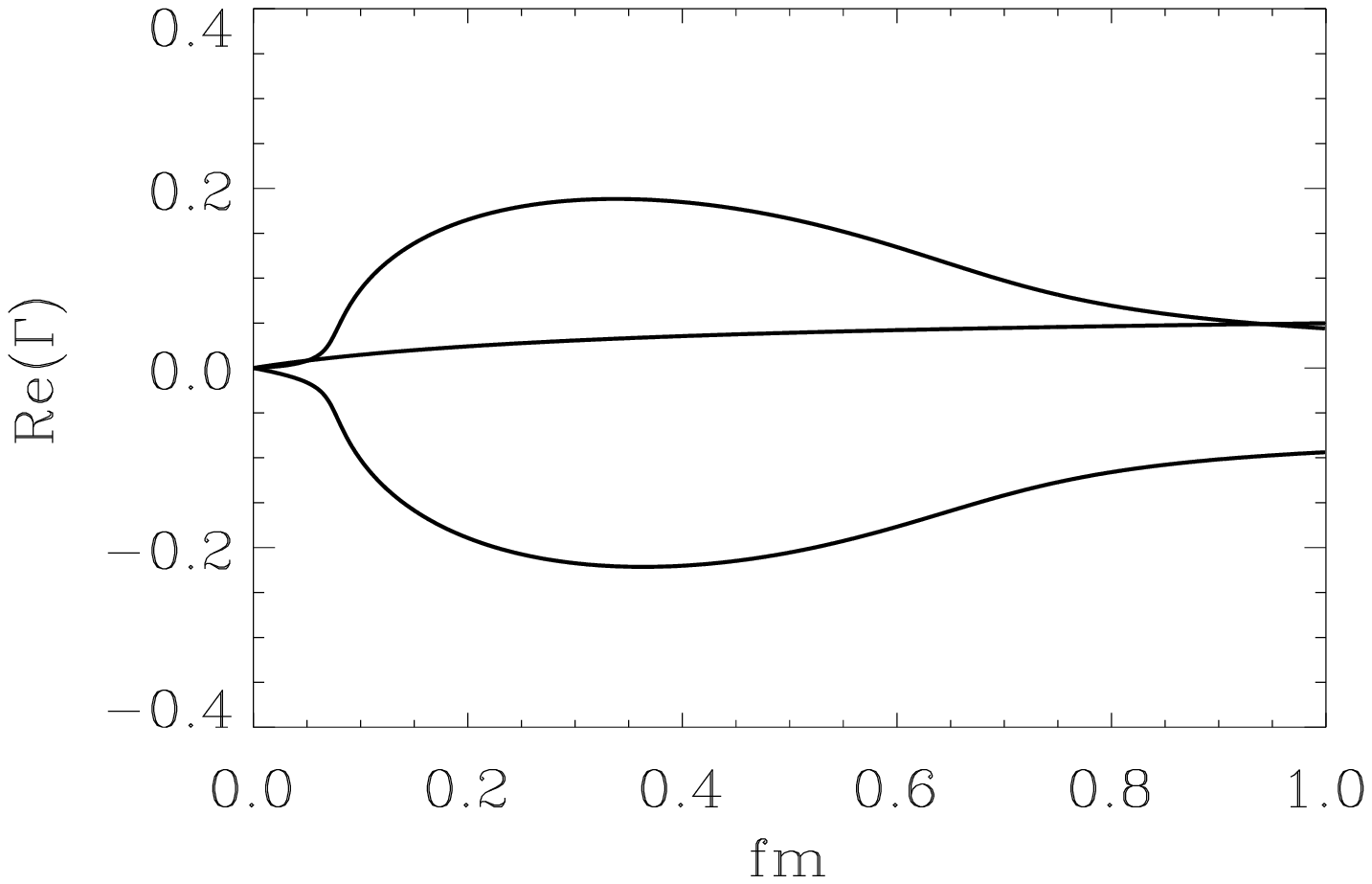}
\caption{Same as in Fig.~4 but for $f_{s}=0.3$ and $h=2$.}
\end{figure}

In Fig.~7, we plot the dependence of the real and imaginary parts of 
$\Gamma$ on $x^{2}$ for the case of radially decreasing $\Omega(s)$ 
($q=4$) and for $\nu=\varepsilon=0.3$, $\delta=0.2$, $h=o.5$, $f_{s}=
0.3$, and $f_{m}=0.2$. The unperturbed field strength is not sufficient 
to suppress the magnetorotational instability. Solid lines show the 
growth rate and frequency for the complex roots, and dashed lines for 
the real roots. The real roots correspond to the magnetorotational 
instability which, in accordance with condition (41), can occur only 
if $x^{2} < 15$ for the given parameters. In the region of shorter 
wavelengths ($x^{2} > 15$), however, the magnetorotational instability 
is forbidden but the newly found instability can be efficient. In this 
region, there are two pairs of unstable complex conjugate roots and a 
pair of stable roots. The new instability is oscillatory, and arises 
despite the magnetic field is rather strong (the thermal and magnetic 
energy are comparable). Even in such a strong field, the growth time of 
instability is rather short, $\sim 3 P$. Note that the growth rate is 
weakly dependent on the wavelength for $x^{2} > 15$. 

\begin{figure}
\plotone{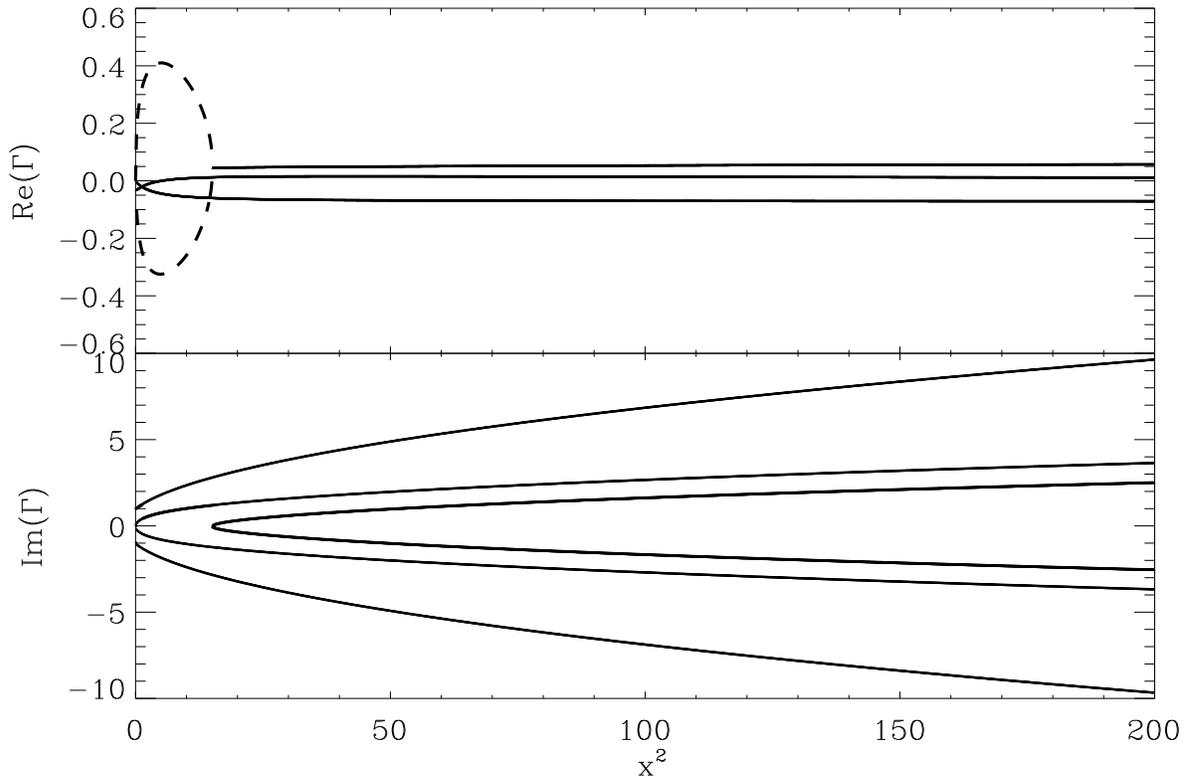}
\caption{The dependence of the growth rate and frequency on $x^{2} = 
(ks)^{2}$ for radially decreasing $\Omega(s)$ ($q=4$) and for $\mu=
\varepsilon=0.3$, $\delta=0.2$, $h=0.5$, $f_{s}=0.3$ and $f_{m}=0.2$.}
\end{figure}

%

In Fig.~8, we plot the same dependence as in Fig.~7 but for the flow with 
$q=0.5$ which is stable to the magnetorotational instability at any 
magnetic field. Only complex modes exist in such a flow and, as a 
result, instability is oscillatory. In this case, Re $\Gamma$ is a bit 
smaller than for a flow with the Keplerian angular velocity ($q=4$) but, 
because of a high epicyclic frequency  $\Omega_{e} = 2 \sqrt{2} \Omega$, 
the growth time is rather short, $\sim P$. As in the previous figure, the 
growth rate is approximately same for all short wavelength perturbations.

\begin{figure}
\plotone{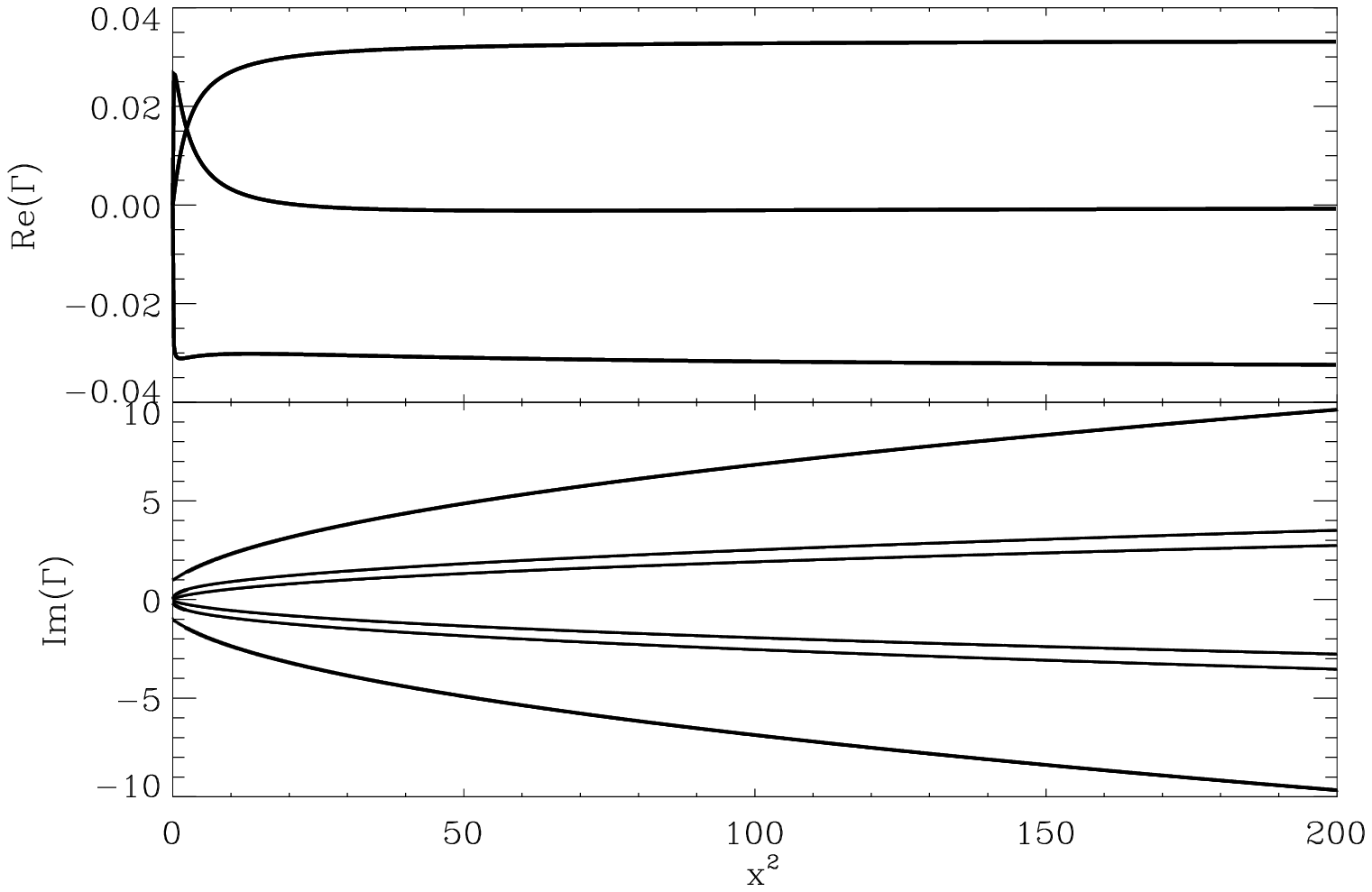}
\caption{Same as in Fig.~7 but for $q=0.5$.}
\end{figure}

Finally, we plot in Fig.~9 the dependence of the growth rate on $x^2$
for rapidly rotating gas with $f_m = f_s = 7 \times 10^{-3}$. This 
figure illustrates the type of solutions that can exist in a relatively
cold plasma. For such solutions, the growth rate is rather high, $\sim 
\omega_{B \Omega}$. Remarkably, that this high growth rate is achieved
for perturbations with $k^2 \sim \Omega^2/ c_A^2$ or, using the dimensionless
parameters, $x^2 \sim 1/ f_m$. The calculated growth rate is in a reasonable
agreement with estimate (26) since $B_s/B_{\phi} \approx 0.1$ in the 
considered case. The growth rate is smaller than it is given by Eq.~(29) for 
very large $x^2$. Note that the roots (27)-(28) were calculated under 
assumption $c_A \gg c_s$ whereas $c_s = c_A$ in the numerical solution.
Therefore, even the number of modes is different in Eqs.~(24)-(25) and in
Fig. 9. However, estimate of the growth rate (29) is still valid despite 
this difference. 

\begin{figure}
\plotone{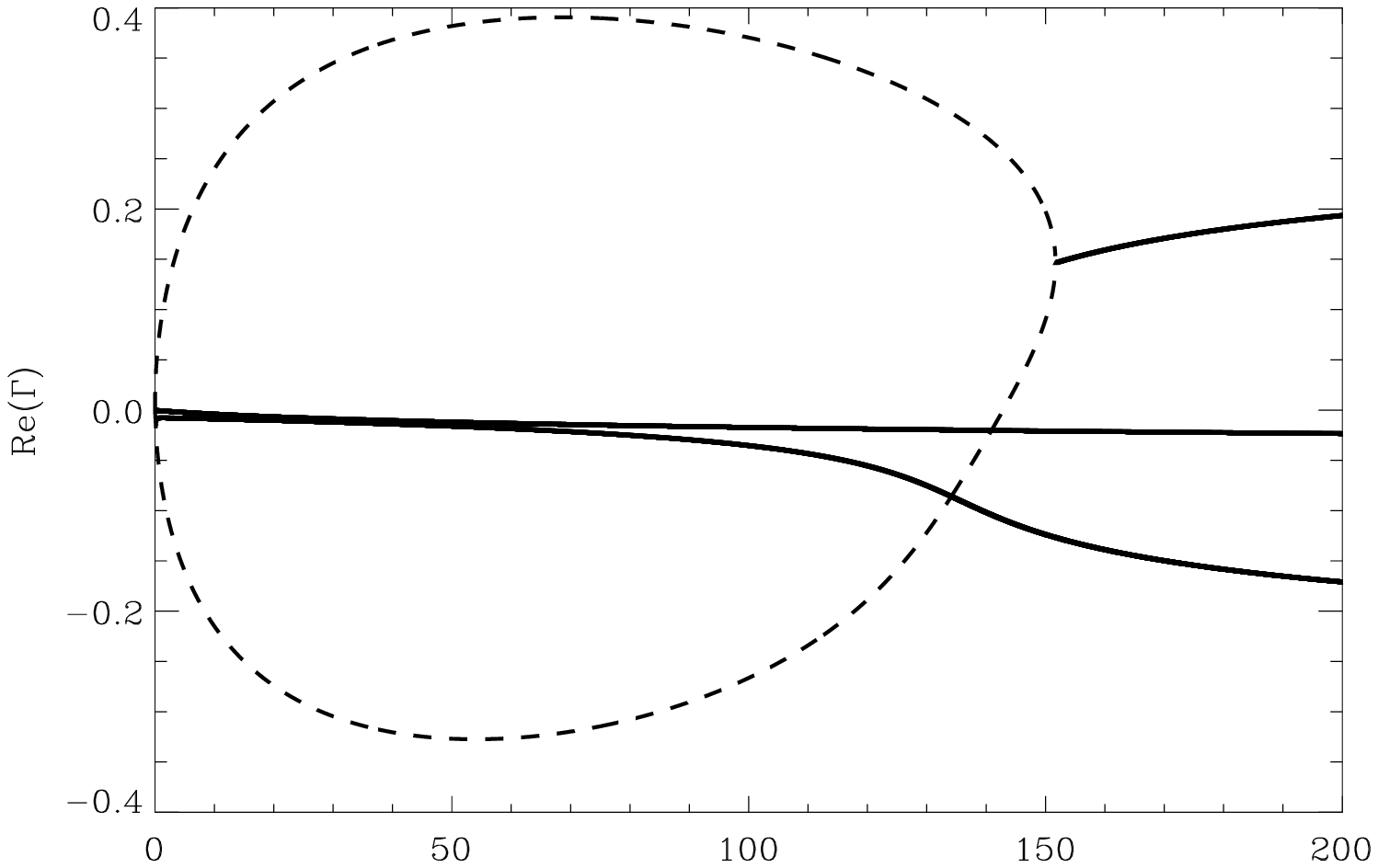}
\caption{The dependence of the growth rate on $x^2$ for $q=4$,
$\mu =\delta= 0.1$, $\varepsilon=0.3$, $h=0.9$, and $f_{s}= f_{m} =7 
\times 10^{-3}$.}
\end{figure}

\section{Discussion}

We have considered stability of differentially rotating magnetized 
gas taking account of the effect of compressibility and assuming that 
the magnetic field has non-vanishing radial and azimuthal components.
In our stability analysis, we assume that the basic state is 
quasi-stationary. This assumption can be fulfilled in many cases of 
astrophysical interest despite the development of the azimuthal field 
from the radial one due to differential rotation. For instance, 
if the magnetic Reynolds number is large then the timescale of 
generation of the toroidal field is $\sim \tau_{\varphi}$ (see 
Eq.~(8)) in the case $B_{\varphi}(0) > B_{s}$. Obviously, the basic 
state can be considered as quasi-stationary if the growth time of 
instability is shorter than $\tau_{\varphi}$. As it is seen from 
Eq.~(29), the growth rate of instability can be roughly estimated as 
$\Omega (B_{s}/B_{\varphi})^{1/3}$ if $s \Omega' \sim \Omega$. Then, the 
condition of quasi-stationarity reads 
\begin{equation}
B_{\varphi} \gg B_{s}.
\end{equation}

The physics of the instability is very simple. To clarify its nature, 
we consider the simplest case of a cold gas with $c_A \gg c_s$. Assume 
that the background magnetic field has only radial $B_s$ and azimuthal 
$B_{\varphi}$ component, and the perturbation is a plane wave in the 
$z$-direction, such as $\vec{k} = (0, 0, k_z)$. The wave produces the 
density perturbation $\rho_1$ that is also an oscillatory function of $z$. 
Since the magnetic field is assumed to be frozen into the gas,
compression/decompression should perturb the magnetic field. The perturbation 
of the radial component is simply related to the density perturbation, $B_{1s}
= B_s (\rho_1/\rho)$. The azimuthal perturbation is determined by two
processes and can be written as $B_{1 \varphi} = B^{(c)}_{1 \varphi} +
B^{(s)}_{1 \varphi}$, where $B^{(c)}_{1 \varphi} = B_{\varphi} (\rho_1 /
\rho)$ is the perturbation caused by compression/decompression in the
wave and $B^{(s)}_{1 \varphi}$ is caused by stretching of the azimuthal
field from the perturbation of the radial component due to differential 
rotation. The rate
of stretching is $s \Omega' B_{1s}$ and, hence, $B^{(s)}_{1 \varphi} = 
s \Omega' \int B_{1s} dt'$. Then, the perturbation of the azimuthal field 
is given by
\begin{equation}
B_{1 \varphi} = B_{\varphi} \frac{\rho_1}{\rho} +
s \Omega' B_s \int \frac{\rho_1}{\rho} dt'. 
\end{equation}
Note that this equation is exactly equivalent to the $\varphi$-component
of Eq.~(16) if we assume $\vec{k} \cdot \vec{B}= 0$ and substitute 
$i (\vec{k} \cdot \vec{v}_{1}) = \dot{\rho}_{1}/\rho$ as it follows from
the continuity equation (14). 
The perturbation of the field leads to the Lorentz force which, in 
its turn, generates the vertical velocity. The Lorentz force is equal
to $(\nabla \times \vec{B}_1) \times \vec{B} / 4 \pi = i \vec{k} (\vec{B}
\cdot \vec{B}_1)/ 4 \pi$. Then, the perturbation of the vertical velocity
is
\begin{equation}
v_{1z} = \frac{i k_z}{4 \pi \rho} \int (\vec{B} \cdot \vec{B}_1) dt'.
\end{equation} 
In its turn, vertical motions should generate perturbations of the 
density since the gas is compressible. We have from the continuity 
condition
\begin{equation}
\frac{\rho_1}{\rho} = i k_z \int v_{1z} d t'.
\end{equation} 
Combining Eqs.~(43)-(45), we obtain the expression governing density
perturbations
\begin{equation}
\frac{\rho_1}{\rho} = - \frac{k_z^2}{4 \pi \rho} \int dt' \! \int dt''
\! \left( \! B^{2} \frac{\rho_1}{\rho} + s \Omega' B_s B_{\varphi} \int
\! \frac{\rho_1}{\rho} dt''' \! \right).
\end{equation}    
If we assume that $\rho_1 \propto \exp(\sigma t)$, then the growth
rate is determined by the condition 
\begin{equation}
\sigma^3 + \sigma \omega_0^2 + \omega_{B \Omega}^3 =0.
\end{equation}
This equation describes the same magnetic modes as Eq.~(24) does in
the case $\vec{k} = k_z \vec{e}_z$. Indeed, in the case $\vec{k} = (0,
0, k_z)$ (or $\mu =1$), Eq.~(24) can be transformed into
\begin{equation}
(\sigma^2 + \Omega_e^2)(\sigma^3 + \sigma \omega_0^2 + 
\omega_{B \Omega}^3) =0,
\end{equation}
such as the modes affected by the magnetic field are described by Eq.~(47).
It is seen from this simple explanation, that shear and compressibility
play the major role in the instability.

In the incompressible limit that applies if $c_{s} \gg c_{A}$, 
differentially rotating flows can be subject to the well-known 
magnetorotational instability (Velikhov 1959, Chandrasekhar 1960) that 
arises if the angular velocity decreases with the cylindrical radius. 
It turns out, however, that compressibility alters drastically the 
stability properties of magnetized rotating flows. Apart from 
instabilities caused by ``the curvature terms'' (see Pessah \& Psaltis
2005), a number of new instabilities can occur in the presence of a radial
magnetic field. The properties of these instabilities are very much 
different from those of the magnetorotational instability. The necessary 
condition of these new 
instabilities is given by Eq.~(35) and reads $B_{s} B_{\phi} \Omega' 
\neq 0$. Since differential rotation always generates the azimuthal 
magnetic field from the radial one, the necessary and sufficient 
condition of instability is the presence of a radial field in 
differentially rotating flows. Obviously, this condition can widely be 
satisfied in astrophysical bodies. Note that, contrary to the 
magnetorotational instability that occurs only if the angular velocity 
decreases with the cylindrical radius, the newly found instabilities may 
occur at any sign of $\Omega'$. 


Likely, the most important difference between the magnetorotational and
newly found instabilities is associated with the dependence on the 
magnetic field strength. A sufficiently strong magnetic field, 
satisfying the inequality $(\vec{k} \cdot \vec{B})^{2} > 8 \pi \rho s 
\Omega |\Omega'| (k^{2}_{z}/k^{2})|$, suppresses  completely the 
standard magnetorotational instability. On the contrary, the instability 
found in our paper cannot be suppressed even in relatively strong magnetic 
fields (but satisfying inequality (23))as it is seen from the criterion 
(35). Often, the growth rate 
reaches some saturation in very strong magnetic field, and this 
saturated value can be rather high (Re $\Gamma \sim 0.1-0.5$).  

The growth rate of the newly found instabilities is rather high and can 
even be comparable to $\sim \Omega_{e}$ for some values of parameters. 
Generally, the growth rate depends on compressibility,  being smaller 
for a low compressibility. The incompressible limit (Boussinesq 
approximation) corresponds to $c_{s} \gg c_{A}$, and the considered 
instabilities are inefficient in this limit because of a low growth 
rate. However, in the case of a strong field with $c_{A} \sim c_{s}$ 
when the Boussinesq approximation does not apply, the instability can 
be much more efficient than the magnetorotational instability.  


\vspace{0.5cm}

\noindent
{\it Acknowledgments.}
This research project has been supported by a Marie Curie Transfer of
Knowledge Fellowship of the European Community's Sixth Framework
Programme under contract number MTKD-CT-002995.
VU thanks also INAF-Ossevatorio Astrofisico di Catania for hospitality.

{}

\end{document}